# Graphene thermal infrared emitters integrated into silicon photonic waveguides

*Nour Negm[1,2], Sarah Zayouna[3], Shayan Parhizkar[1,2], Pen-Sheng Lin[4], Po-Han Huang[4], Stephan Suckow[1], Stephan Schroeder[3], Eleonora De Luca[3], Floria Ottonello Briano[3], Arne Quellmalz[4], Georg S. Duesberg[5], Frank Niklaus[4], Kristinn B. Gylfason[4], Max C. Lemme[1,2]*

[1]AMO GmbH, Advanced Microelectronic Center Aachen, Otto-Blumenthal-Str. 25, 52074 Aachen, Germany

[2]Chair of Electronic Devices (ELD), RWTH Aachen University, Otto-Blumenthal-Str. 25, 52074 Aachen, Germany

[3]Senseair AB, Stationsgatan 12, 824 08 Delsbo, Sweden

[4]Division of Micro- and Nanosystems, School of Electrical Engineering and Computer Science, KTH Royal Institute of Technology, 100 44 Stockholm, Sweden

[5]Institute of Physics, Faculty of Electrical Engineering and Information Technology (EIT 2), Bundeswehr University Munich, Werner-Heisenberg-Weg 39, 85577 Neubiberg, Germany

ABSTRACT

Cost-efficient and easily integrable broadband mid-infrared (mid-IR) sources would significantly enhance the application space of photonic integrated circuits (PICs). Thermal incandescent sources are superior to other common mid-IR emitters based on semiconductor materials in terms of PIC compatibility, manufacturing costs, and bandwidth. Ideal thermal emitters would radiate directly into the desired modes of the PIC waveguides via near-field coupling and would be stable at very high temperatures. Graphene is a semi-metallic two-dimensional material with comparable emissivity to thin metallic thermal emitters. It allows maximum coupling into waveguides by placing it directly into their evanescent fields. Here, we demonstrate graphene mid-IR emitters integrated with photonic waveguides that couple directly into the fundamental mode of silicon waveguides designed for a wavelength of 4,2 µm relevant for $CO_2$ sensing. High broadband emission intensity is observed at the waveguide-integrated graphene emitter. The emission at the output grating couplers confirms successful coupling into the waveguide mode. Thermal simulations predict emitter temperatures up to 1000°C, where the blackbody radiation covers the mid-IR region. A coupling efficiency η, defined as the light emitted into the waveguide divided by the total emission, of up to 68% is estimated, superior to data published for other waveguide-integrated emitters.

MAIN

The demand for distributed, networked, and compact gas sensors[1–4] for real-time air quality monitoring is rapidly increasing in a wide range of applications such as demand-controlled ventilation in heating and air conditioning (HVAC) systems[5,6], gas leak detection[7], industrial process control[8,9], environmental monitoring[10–12] and medical diagnostics[13]. Typical gas sensing methods are often based on chemical reaction sensors such as catalytic beads or electrochemical and semiconducting metal oxides[1,4]. In these types of sensors, the targeted gas undergoes a chemical reaction with the sensor material and thus alters the sensor itself. This leads to drift over time, the need for frequent calibration, performance degradation, and a limited sensor lifetime. Absorption spectroscopy, in contrast, exploits the fundamental absorption lines of trace and greenhouse gases in the mid-infrared (mid-IR) region. This allows "fingerprinting" the gases through characteristic absorption wavelengths between 3 µm and 10 µm, e.g., carbon dioxide ($CO_2$) at 4.2 µm[14]. Thus, absorption spectroscopy is an optical gas sensing technology that provides high specificity, minimal drift, and long-term stability without chemically altering the sensor[2,3].

Photonic integrated circuits (PICs) enable shrinking spectroscopy equipment to the size of a chip, resulting in highly miniaturized and cost-efficient optical gas sensor systems. However, recent work demonstrating PIC-based gas sensors still requires light coupling from external sources and to detectors into and out of the waveguides, respectively[15–17]. Integrating these components directly on the wafer level would reduce the sensor system size and cost, enhance mechanical stability, and potentially enhance performance, thus enabling the widespread use of such systems as air quality monitors. There is promising progress for waveguide-integrated mid-IR photodetectors (including graphene and other 2D materials)[18–25], but the realization of waveguide-integrated mid-IR emitters has seen very limited success to date. Commonly used mid-IR emitting materials such as small bandgap III-V or II-VI semiconductor compounds cannot easily be integrated with CMOS and PIC substrates because of their fabrication processes and high thermal budgets. They also suffer from high manufacturing costs due to their complex epitaxial multi-layer systems, and their light emission has a limited bandwidth, making them unsuitable for compact multi-gas sensor applications requiring broadband emission[26–30].

Alternatively, thermal incandescent light sources are well-established and cost-effective mid-IR sources. They work by Joule heating realized by forcing an electric current through the emitter material, causing broadband thermal black body emission according to Planck's law. The precise spectral distribution of the emitted radiation depends mainly on the temperature of the emitter, but it is, in any case, broadband. The design of thermal incandescent sources is comparably simple, as they are generally composed of a layer of an electrically conductive emitting material and two electrical contact pads. Such emitters can be integrated on-chip with optoelectronic components and PICs in high-volume production flows. The latter enables near-field coupling of their emission directly into the waveguided modes. Graphene is an excellent candidate for such emitters, as it has been shown to reach the required temperatures for thermal emission in the mid-IR[31]. Its emissivity is also comparable to that of other very thin emitters[32,33]. At the same time, monolayers of graphene are so thin that the entire emitting volume can be placed closest to the waveguide, generating ideal near-field coupling. Thicker metallic emitters partly shield their own radiation. Finally, monolayer graphene only minimally distorts the waveguided mode, which minimizes the mismatch between the mode in the emitter region and outside this region.

In this work, we show a waveguide-integrated incandescent thermal mid-IR emitter that uses graphene as the active material. We experimentally demonstrate graphene emitters integrated directly on top of silicon photonic waveguides. The emitters couple directly into the waveguide mode. We detect emissions in the spectral range of 3 µm to 5 µm, including our target of 4.2 µm for the detection of $CO_2$.

RESULTS

**Device layout and waveguide simulation**

A sketch of the cross-section, a schematic top view, and a scanning electron microscope (SEM) image of our waveguide-integrated emitters are shown in Figure 1 a-c). The thermal emitter mainly consists of a graphene sheet on top of the Si waveguide, with contacts on each side. A layer of $Al_2O_3$ was used to encapsulate the graphene from the environment (Methods). Two grating couplers are included to outcouple radiation with a wavelength of $\lambda = 4.2$ µm at each end of the waveguides. The fundamental quasi-TE mode profile in the waveguides calculated for $\lambda = 4.2$ µm (Methods) is shown in Figure 1 d).

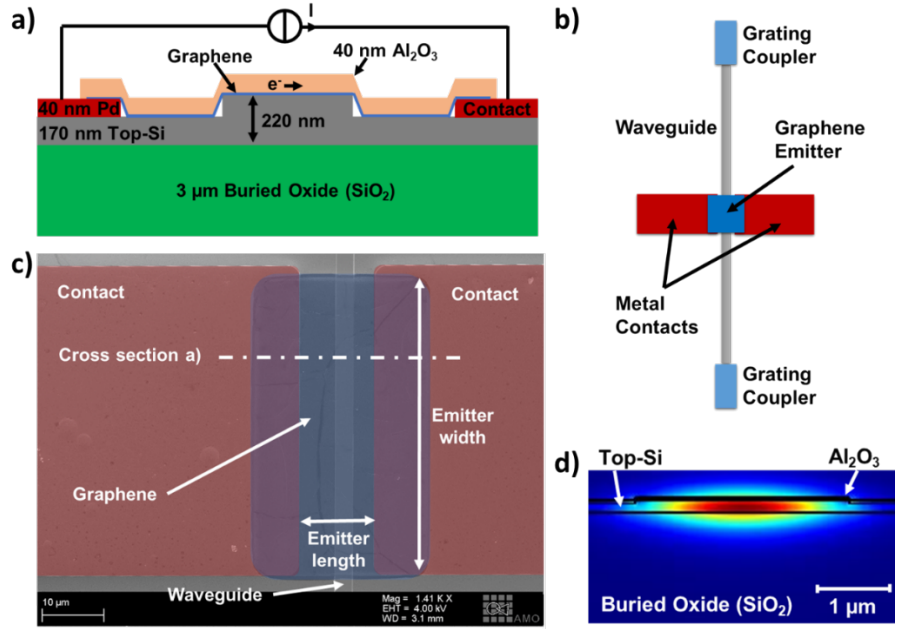

**Figure 1. Device layout and waveguide simulation. a)** Schematic cross-section of the graphene waveguide-integrated thermal emitter after encapsulation. The electrical connection scheme illustrates the Joule heating method to generate thermal emissions. **b)** Schematic top view of the components of the photonic integrated circuit. **c)** False-colored scanning electron micrograph before encapsulation. **d)** Simulated electric field profile of the fundamental TE mode of the waveguide, including relevant materials.

**Mid-IR setup and measurements**

Two probe needles were used to contact the graphene emitters via the metal pads with a source measure unit (SMU), which acted as a DC current source and voltage meter simultaneously (Figure 2 a)). A DC current was forced into the emitters starting at 5 mA and ramped up to 25 mA in incremental steps of 0.2 mA for 5 s, and the voltage drop across the emitters was monitored. In addition, we captured top-view IR images of the devices with an infrared camera with a spectral wavelength range from 3 µm to 5 µm. The camera was positioned above the chip to detect the light emitted at the emitter and one of the two grating couplers at the end of the waveguide. All measurements were conducted in an ambient atmosphere with 10% to 20% humidity at 22°C.

An IR image of a waveguide-integrated graphene emitter operated at a current density of 0.45 mA/µm (current normalized to graphene channel width), corresponding to a power of 185 mW, is shown in Figure **2** b). Mid-IR emission is visible both at the graphene emitter and at the output grating coupler. The emission at the grating couplers clearly demonstrates that the graphene emits IR radiation, which is then coupled into and propagates through the waveguide to the grating couplers. This proves the functionality of the graphene emitter and its successful integration with the photonic waveguide. The sequence of images in Figure **2** c) illustrates the power dependency of the emission. Note that additional emission visible at the probe tips is due to reflections of the emitted light. This is therefore disregarded in the following analysis.

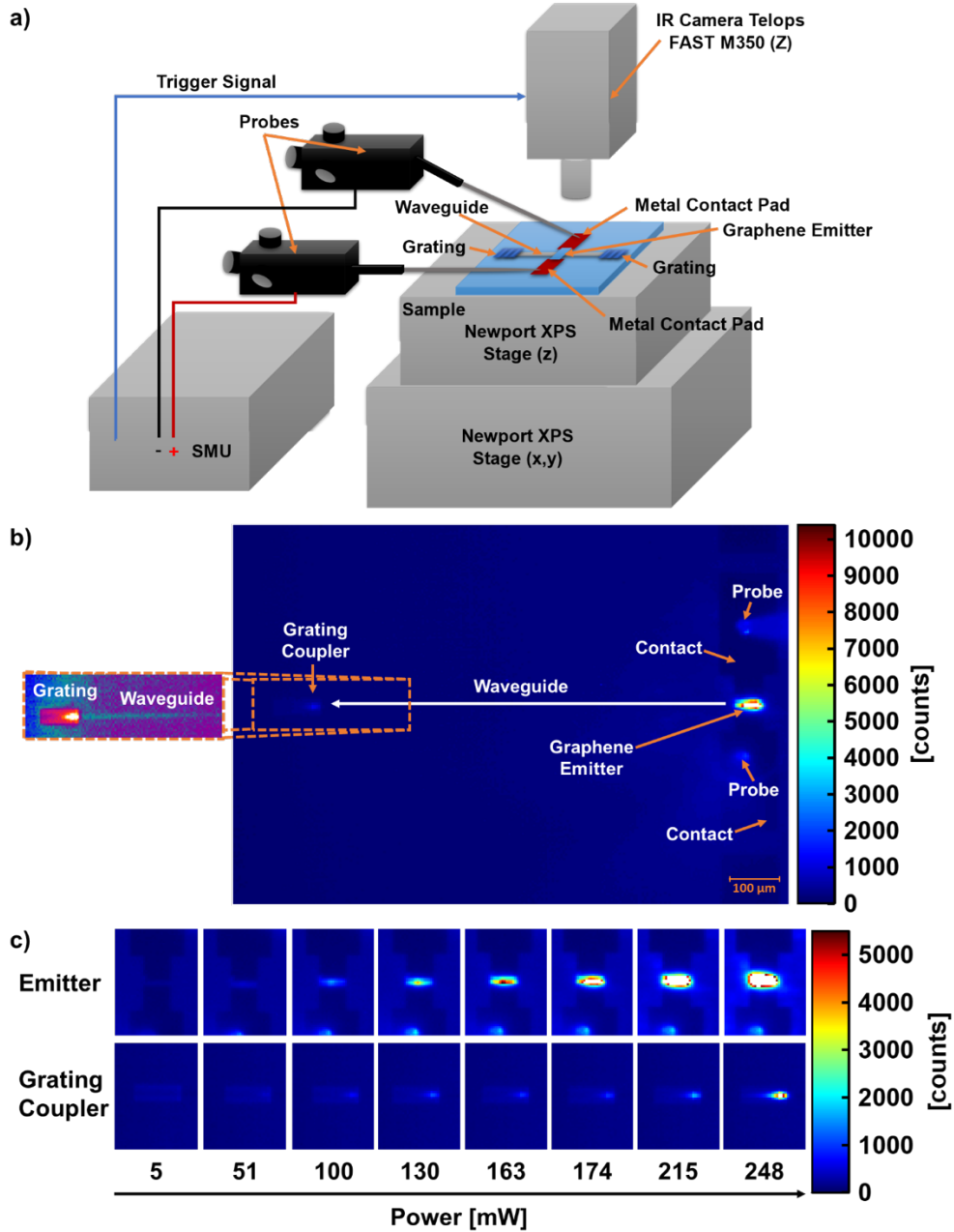

**Figure 2. Mid-IR setup and measurements. a)** Schematic of the mid-IR measurement setup, showing the components and connections to the integrated thermal emitter on the sample. In b) and c) the camera focuses on the emitter and one of the output grating couplers. **b)** Mid-IR camera image of a thermal emitter operated at 185 mW. The emitter, contact pads, and probes are visible on the right. The waveguide output grating coupler is visible on the left. The length of the waveguide between the grating coupler and emitter is 725 μm. **Inset:** Extraction of the output

grating with enhanced contrast (using a different color scale), to visualize the radiation pattern of the outcoupled light and the scattering from the waveguide. The scale bar is valid for the entire Figure b). **c)** Mid-IR camera images of the emitter and grating coupler from b), at different power levels.

The emission intensity at the output grating coupler is significantly lower than at the emitter. There are several reasons for this. Firstly, the graphene emitter exhibits a typical gray body broadband emission, which the IR camera partly captures in the wavelength range from 3 µm to 5 µm. However, the grating couplers are wavelength-selective and filter out some of the broadband radiation coming from the emitter. Even for the optimal wavelength, the outcoupling efficiency is far below unity (details in supplementary Information SM1). Secondly, the emitter-waveguide coupling is not perfect and leads to losses, as well as waveguide propagation losses. We measured the waveguide propagation losses at $\lambda = 4.2$ µm with the cutback method and obtained $14.6 \pm 3.5$ dB/cm (details in supplementary Information SM1). This value is close to the intrinsic waveguide propagation loss of 13.7 dB/cm, calculated with the Ansys Lumerical Mode Solver. As expected, the simulation underestimates the loss because the model only includes absorption in the BOx layer but not scattering losses. The device shown in Figure **2** with its length of 726 µm thus contributes a total waveguide propagation loss of approximately 1 dB.

**Emitter coupling efficiency**

We estimated the coupling efficiency $\eta$ of the emitter from the camera images as shown in Figure **2**. $\eta$ is defined as the intensity of the light emitted from the emitter into the waveguide $I_{E,in}$, divided by the total emissions of the emitter, $\eta = I_{E,in}/(I_{E,out} + I_{E,in})$, with $I_{E,out}$ being the emission intensity extracted above the emitter by the camera. First, the background illumination level was subtracted in each image using an off-state IR image of the area under investigation. Then, the emission intensity at the grating coupler $I_{GC,out}$ and at the emitter $I_{E,out}$ was extracted by integrating the camera counts around the respective emission peak positions. Note that we are using counts for convenience, but this is not raw data but data calibrated for linearity. We defined the emission area as the set of pixels with more counts than 1/e of the peak value. The emission intensity was $I_{GC,out}$ = 9,154 counts at the grating coupler and $I_{E,out}$ = 230,990 counts at the emitter based on the image in Figure **2** b). Although this methodology excludes photons interacting with the surface or emitted at angles outside of the numerical aperture of the camera, we expect this data to be at least sufficient for calculating an apparent coupling efficiency. Next, to calculate $I_{E,in}$ from $I_{GC,out}$, we considered the optical losses of the waveguide and the grating coupler. We used the measured waveguide loss of 1 dB, assuming that the propagation loss measured at 4.2 µm is representative of the entire wavelength range captured by the camera (3 – 5 µm). Making the same assumption for the optical losses of the grating coupler is not possible due to the strong spectral filtering of the grating. We, therefore, extended the simulation of the grating coupler efficiency to include the camera's sensitivity range and obtained an average value of approximately 23%. We further assumed symmetrical emission into each side of the waveguide. The emission into the waveguide can then be calculated to $I_{E,in} = 2I_{GC,out}/0.8/0.23 = 99,500$ counts, and the efficiency of the emitter was calculated to be $\eta$ = 30%.

The measured emission intensities of the graphene emitter and output grating coupler and the coupling efficiency of the emitter are plotted as a function of electrical power in Figure 3. The intensities at the grating coupler and emitter start to distinctly stand out from the background at a power of approximately 40 mW to 50 mW, in line with the camera images in Figure **2** c). The efficiency calculation should therefore be disregarded below this value, as the data widely fluctuates (grayed out). Over 50 mW, the efficiency is between 55% and 65%, with a maximum of 68% at 123 mW. It then decreases to around 40% in the power range of 150 mW to 174 mW and around 30% as the emitter starts exhibiting signs of a thermal runaway at 185 mW. The efficiency increases again to 40% and then over 80% just as the device breaks down.

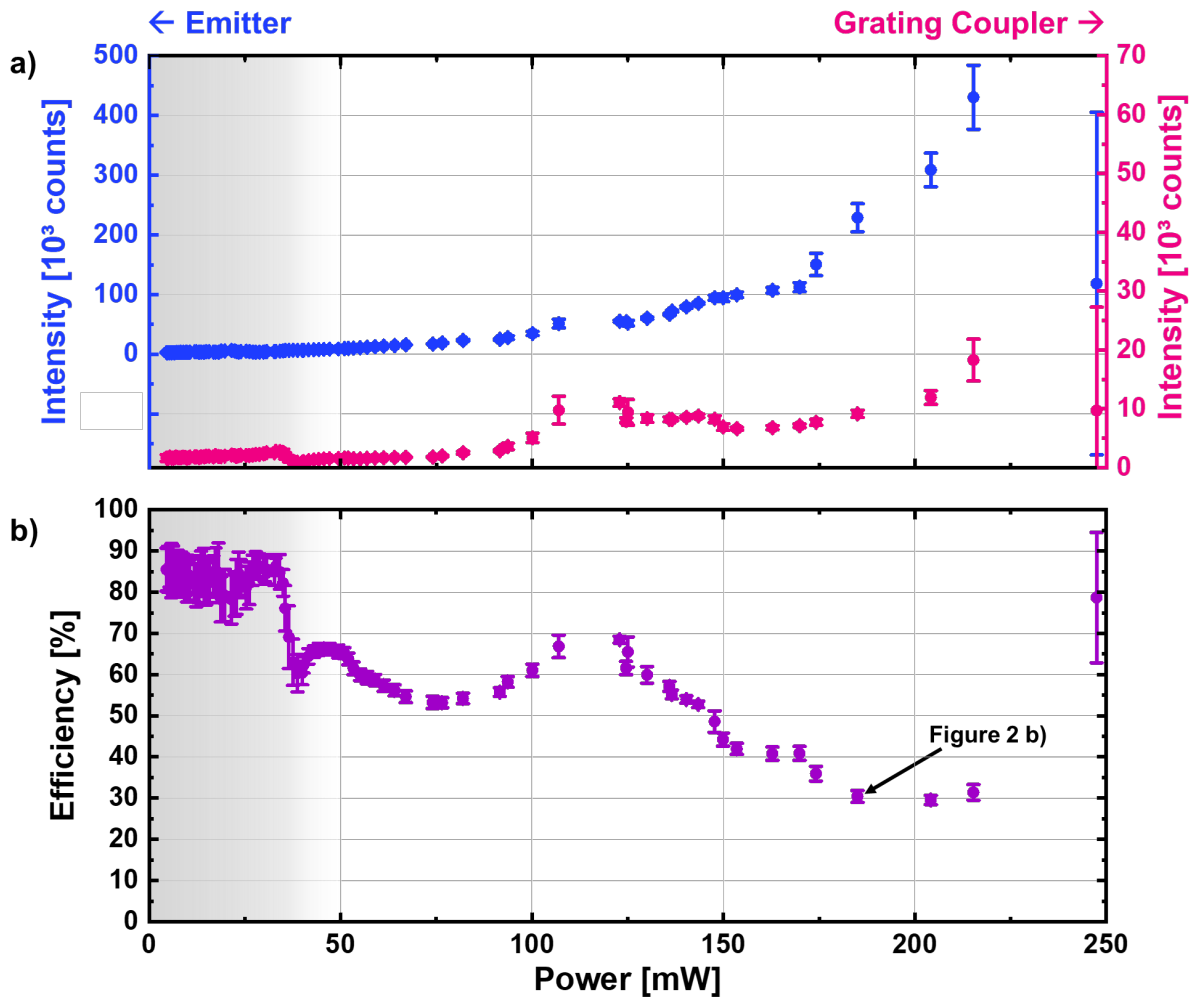

**Figure 3. Emission efficiency.** *Version3-4* **a)** Mean and standard deviation of the emission intensities of the graphene emitter and output grating coupler as a function of DC electric power extracted from IR images as illustrated in Figure **2**. The IR camera recorded 50 frames per current

increment. The two data sets are offset for clarity. **b)** Calculated coupling efficiency of the graphene emitter (mean and standard deviation) as a function of electric power.

Wien's displacement law offers a qualitative explanation of the measured efficiency decrease with heating power, i.e., emitter temperature. The law states that high temperatures shift the peak emission of the thermal emitter to shorter wavelengths. The free space emission at the emitter side, however, is not affected by this spectral change. The camera captures all radiation in the waveband of 3 to 5 µm within the numerical aperture of the objective, which translates to an angle of 26.5°. However, the emission at the grating coupler is expected to vary from an angle of -20° at 5 µm wavelength to 40° at 3 µm wavelength. Therefore, the shortest wavelength emission is not collected by the objective. The fraction of emission at 3 µm wavelength increases with increasing emitter temperature, leading to an apparent drop in coupling efficiency measured by our camera.

**Operating regimes and stability**

We further explored the potential operating regimes and stability of the emitters. Figure **4** a) shows the emitter emission as a function of DC electric power for seven nominally identical emitters. Two operating regimes can be identified. There is stable operation below a power threshold of approximately 200 mW and unstable operation characterized by fluctuations and thermal runaway between 250 mW and 350 mW before final breakdown at different power levels.

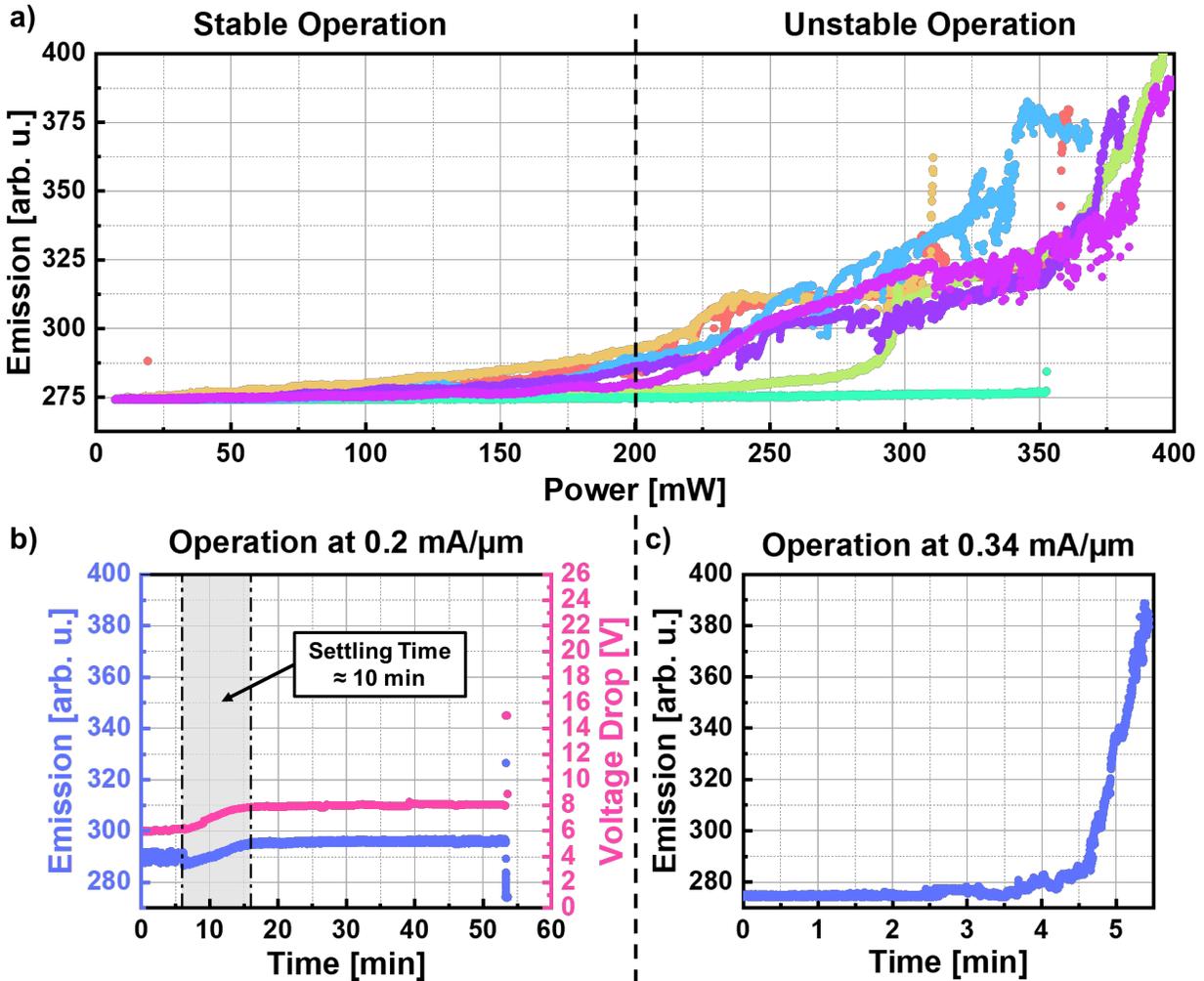

**Figure 4. Operating regimes and stability. a)** Emission in arbitrary units (arb. u.) – as extracted from the IR camera software – of seven nominally identical emitters versus input power. **b)** Emission in arbitrary units and voltage drop of an emitter driven in the stable regime with 0.2 mA/µm, corresponding to 60 – 80 mW, versus time. **c)** Emission in arbitrary units of an emitter driven in the unstable regime with 0.34 mA/µm versus time.

Figure **4** b) shows the emission of and the voltage drop along one device operated in the stable regime with a constant current of 0.2 mA/µm over time. The device was operated continuously for a total of 54 min. After an initial stable plateau for the first 6 min, we observed an increase in emission and voltage over 10 min. The voltage change originates from a resistance change of the emitter, which may be attributed to current-induced damage to parts of the graphene sheet or the graphene-metal contacts, which increased the total resistance of the emitter and thus the heating power (here: from 60 mW to 80 mW). Afterwards, the device was stable for 40 min before the measurements were terminated. Another device was operated at a current density of 0.34 mA/µm. Although the voltage drop across this device was not measured, we consider it to be in the unstable operating regime based on typical average resistances of the graphene sheets. This device remained stable for only 2.5 min, after which the emission started to increase in a thermal runaway until the device broke down after 5.4 min. The thermal runaway is expected to be initiated by local current-induced defects in the graphene, which increase the resistance and, as the supplied current is constant, increase the heating power. The resulting enhanced emission comes at the price of defect generation and, ultimately, device failure.

**Electro-thermal simulation**

The electro-thermal behavior of the emitters was further investigated with finite element simulations using COMSOL Multiphysics software (see Methods section). Two different device configurations were modeled, (1) one where the graphene sheet is suspended and only supported by the metal pads and the silicon waveguide, and (2) one where the graphene adheres completely to the topography of the metal pads, the silicon waveguide, and the substrate (see schematics in Figure 5 a)). In the actual devices, the graphene sheet is assumed to be in a configuration between those two model configurations (compare Figure 1 a)). The simulated temperature distributions of the emitters at a power supply of 185 mW, which corresponds to the experimental data in Figure **2** b), are shown in Figure 5 b). The maximum temperature of the suspended graphene sheet is 803 K, while the graphene that is supported by the Si reaches a temperature of 629 K. The maximum value of the former is located at the centers of the two suspended regions, where the cooling by the environment is the weakest. The fully supported graphene is cooled more efficiently through contact with the substrate and reaches its maximum temperature at the center of the waveguide. The two "hot spots" of the suspended emitter make this configuration more efficient as a light source, as the power emitted by a black body scales with the fourth power of the temperature. However, this configuration also increases the risk of thermal runaway and breakdown, which mostly happened in our case near the contact edges and the edges of the waveguides (see supplementary information SM2).

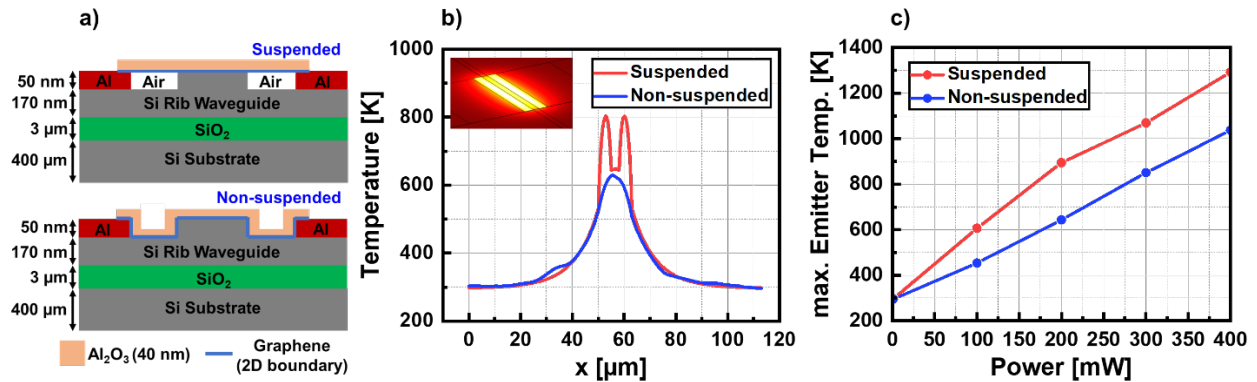

**Figure 5. Electro-thermal simulation. a)** Schematic of the two electro-thermal model configurations used to predict the temperature of the graphene emitters. **b)** Simulated temperature distribution of the graphene sheets versus position perpendicular to the waveguide at 185 mW power supply. **c)** Simulated maximum temperature of the graphene emitters versus power.

The simulations in Figure 5 c) show that both emitter types reach temperatures well over 500 K in the stable experimental operating regime. This corresponds to blackbody radiation that sufficiently covers the mid-IR region. These theoretical results confirm that our graphene waveguide-integrated thermal emitters are suitable mid-IR sources.

DISCUSSION

Table 1 compares the performance of our graphene emitters with the current state of the art achieved with both conventional materials and other nano-emitters. Our devices reach comparable temperatures, which is the most important parameter for thermal emitters. Our graphene emitters further have advantages over the metallic emitters in terms of power consumption and emission area, as they can be directly integrated on top of waveguides and in larger areas without the typical trade-off with reabsorption losses caused by the emitting material (assuming perfect transparency of graphene). The estimated coupling efficiency in the range of 28% to 68% of our graphene emitters compares well with the published CNT emitter[34], even if optical propagation losses and scattering of other wavelengths from the grating coupler were neglected there, which could increase the estimated efficiency by several percent. Nevertheless, graphene emitters would still be favorable, considering the more challenging fabrication processes needed for CNTs, especially on wafer-scale[35].

**Table 1.** Comparison of the present work (highlighted in green) with the state-of-the-art. The most relevant values used for comparison are highlighted in yellow. Simulated values are marked by *. CNT = carbon nanotube.

| Ref. | Material | Coupled to | Est. Coupling Efficiency | Current / Current Density | Voltage | Power Cons. / Power Density | Emitting Area | Temperature | Lifetime |
|---|---|---|---|---|---|---|---|---|---|
| **Here** | Graphene monolayer | Waveguide | 28 – 68% | 10 – 27 mA 0.2 – 0.45 mA/µm | 6 – 18 V | 600 – 200 mW 0.4 – 1.2 mW/µm² | 50 x 10 µm² | 500 – 900 K* | > 54 min |
| 36,37 | Kanthal (FeCrAl) | Free space | | 150 – 200 mA | 6.5 V | 0.8 – 1 W 0.8 – 1 W/mm² | 1 x 1 mm² | 700 – 1200 K | |
| 36,38 | Pt Nanowire | Free space | | | | 5 mW 2.9 mW/µm² | 3.5 µm x 500 nm | 595 K | |
| 31 | Graphene | Free space | | 1 – 2 x10⁸ A/cm² | 16 – 21 V | ~ 80 kW/cm² | | 1600 K | |
| 32 | Graphene multilayer | Free space | | | 8 V | | | 750 K* | > 1000 h |
| 34 | CNT (3 – 100 µm⁻¹) | Waveguide | 45 – 55% | | 1.5 – 10 $V_{DC}$ & additional 2 – 3.3 $V_{pulse}$ | | | 1000 – 1500 K* | |

We have successfully integrated graphene thermal emitters on photonic waveguides and demonstrated their mid-IR emission into the waveguide and out of a grating coupler designed for a wavelength of 4.2 µm. Simulations predict emitter temperatures in the range of 500 – 900 K, matching other nano-scale emitter types. We estimate emission coupling efficiencies into the waveguide of up to 68% based on our data, which is favorable over other nano-scale emitters. We have shown operation for at least ~1 h under ambient conditions. While the PIC in our experiment was designed for 4.2 µm wavelength targeting $CO_2$ detection, the integrated graphene thermal emitters radiate in a broad gas absorption "fingerprint" wavelength range of 3 – 10 µm. This, combined with integrated graphene mid-IR photodetectors, makes them ideal for fully integrated photonic sensors, where the evanescent fields of the waveguides can interact directly with the gaseous environment, enabling ubiquitous gas and environmental sensing.

SUPPLEMENTARY INFORMATION

The following files are available free of charge.

SM1 – Experimental loss extraction using the cutback method (DOC).

SM2 – Investigation of the possible breakdown mechanisms (DOC).

ACKNOWLEDGMENTS

The authors thank Daniel Neumaier (Bergische Universität Wuppertal) for fruitful discussions. This work has received funding from the European Union's Horizon 2020 research and innovation programme under grant agreements No 825272 (ULISSES) and No 101017186 (AEOLUS).

METHODS

**Device Fabrication**

The PICs were fabricated from silicon-on-insulator (SOI) substrates with a buried oxide (BOx) layer thickness of $t_{BOx} = 3$ µm and a top silicon (top-Si) thickness of $t_{Top-Si} = 220$ nm. First, 3 µm wide rib waveguides with 50 nm step height were etched into the top-Si by reactive ion etching (RIE), resulting in 170 nm rib height. Two grating couplers per waveguide were also included, designed for outcoupling radiation of a wavelength of λ = 4.2 µm. 40 nm palladium (Pd) contacts were then evaporated and patterned by lift-off. Next, a monolayer of graphene grown by chemical vapor deposition (CVD) on a copper (Cu) foil was wet-transferred onto the chips using a PMMA-assisted wet-transfer method and patterned with an oxygen ($O_2$) plasma RIE process.

The shallow step height of the rib-waveguide allowed the safe graphene transfer without risking mechanical damage at the steep waveguide edges, which feature ~88° sidewall angles. Finally, a 40 nm aluminum oxide layer ($Al_2O_3$) was deposited using atomic layer deposition (ALD) to encapsulate the graphene and patterned by wet-chemical etching.

**Waveguide Simulation**

The fundamental quasi-TE mode profile in the waveguides was calculated for $\lambda = 4.2$ µm via the Ansys Lumerical mode solver. The refractive index of Si was taken from the built-in model, while the index of the ALD-deposited $Al_2O_3$ was measured by spectroscopic ellipsometry. The perfectly matched layer (PML) boundaries were placed sufficiently far to ensure they would not influence the mode profile.

**Electro-thermal simulation**

COMSOL Multiphysics software was used to investigate the electro-thermal behavior of our emitters with finite element simulations. In both device configurations described in the main text, the graphene sheets were modeled using 2D boundaries with an effective thickness of 1 nm, optimizing the computation time. The substrate temperature was set by defining a fixed temperature boundary at 293 K to the bottom surface of the substrate. Air convection was included in the simulations using a typical 5 W/m²K free convective heat transfer coefficient on the top surfaces of the emitters. In the air gaps underneath the suspended graphene, only the air conduction effect was considered since the air gap depth was sufficiently small to neglect the air convection effect. The electrical conductivity of the graphene was set to $6.5 \times 10^5$ S/m, based on an average measured resistance of 400 Ω.

# Graphene thermal infrared emitters integrated into silicon photonic waveguides – Supplementary Information

*Nour Negm[1,2], Sarah Zayouna[3], Shayan Parhizkar[1,2], Pen-Sheng Lin[4], Po-Han Huang[4], Stephan Suckow[1], Stephan Schroeder[3], Eleonora De Luca[3], Floria Ottonello Briano[3], Arne Quellmalz[4], Georg S. Duesberg[5], Frank Niklaus[4], Kristinn B. Gylfason[4], Max C. Lemme[1,2]*

[1]AMO GmbH, Advanced Microelectronic Center Aachen, Otto-Blumenthal-Str. 25, 52074 Aachen, Germany

[2]Chair of Electronic Devices (ELD), RWTH Aachen University, Otto-Blumenthal-Str. 25, 52074 Aachen, Germany

[3]Senseair AB, Stationsgatan 12, 824 08 Delsbo, Sweden

[4]Division of Micro- and Nanosystems, School of Electrical Engineering and Computer Science, KTH Royal Institute of Technology, 100 44 Stockholm, Sweden

[5]Institute of Physics, Faculty of Electrical Engineering and Information Technology (EIT 2), Bundeswehr University Munich, Werner-Heisenberg-Weg 39, 85577 Neubiberg, Germany

## SM1 – IR CAMERA IMAGES

Figure S6 illustrates the WG propagation and GC coupling losses extracted employing the so-called cutback measurement method. For that, dedicated test structures of different waveguide lengths between 1.1 mm and 4 mm, with grating couplers on each side, are characterized using an external quantum cascade laser emitting at 4.23 µm wavelength and a suitable single-pixel detector in a fiber-coupled setup.

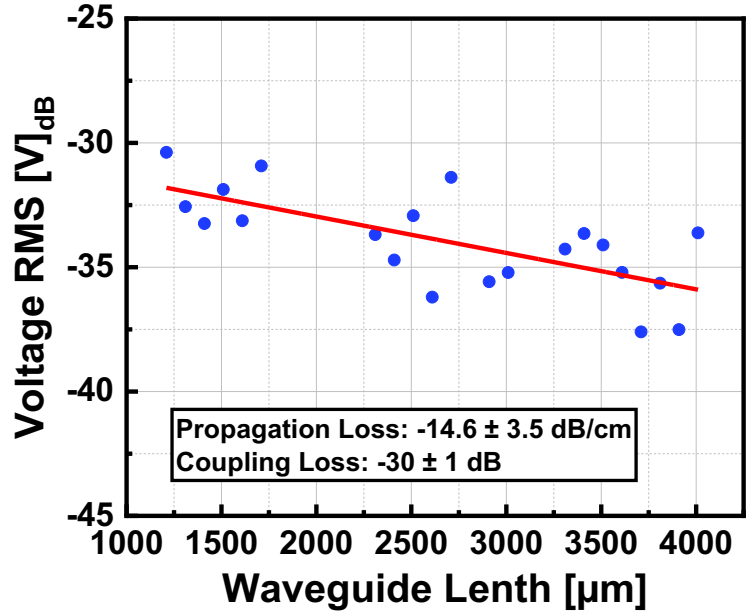

**Figure S6.** Cut-back method result measured at $\lambda = 4.2$ µm, plotting the RMS voltage at the detector in dB versus the length of different test waveguides to determine the propagation loss (slope) and total coupling loss (y-intercept).

SM2 – BREAKDOWN MECHANISMS

The breakdown mechanisms of the devices were investigated *post-mortem* through SEM images (Figure S7). Two breakdown areas have been identified commonly: the contact edges and the edges of the rib waveguides. The former indicates a high resistance of our bottom contacts to graphene. The latter can be attributed to the hot spots identified in the electro-thermal simulation, which may also be present in the actual sample to some extent. Additional defects or stress at the 50 nm step of the WG may further increase the resistance locally and hence lead to higher heating power and earlier breakdown. Some SEM images before breakdown also show other defects in the graphene, like folds in the graphene sheet or residues/particles (Figure S8). Some of these can be expected to create additional local hotspots. The simulation does not include such defects and hence probably underestimates the maximum temperatures reached. It is thus plausible that material starts to melt, as the SEM images in Figure S7 suggest, despite "only" 1000 K being reached in the simulation.

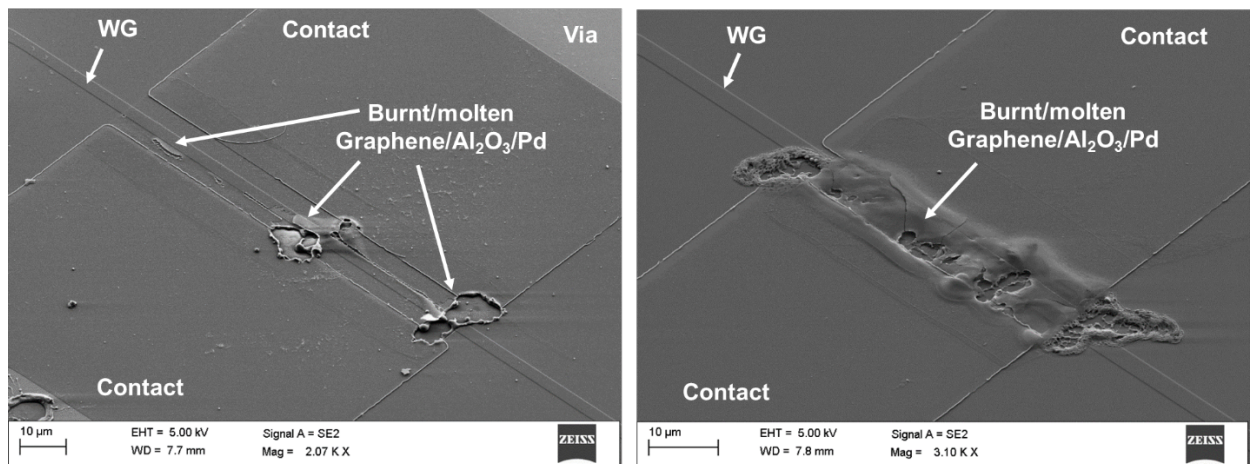

**Figure S7.** Post-mortem SEM images showing the devices after thermal breakdown.

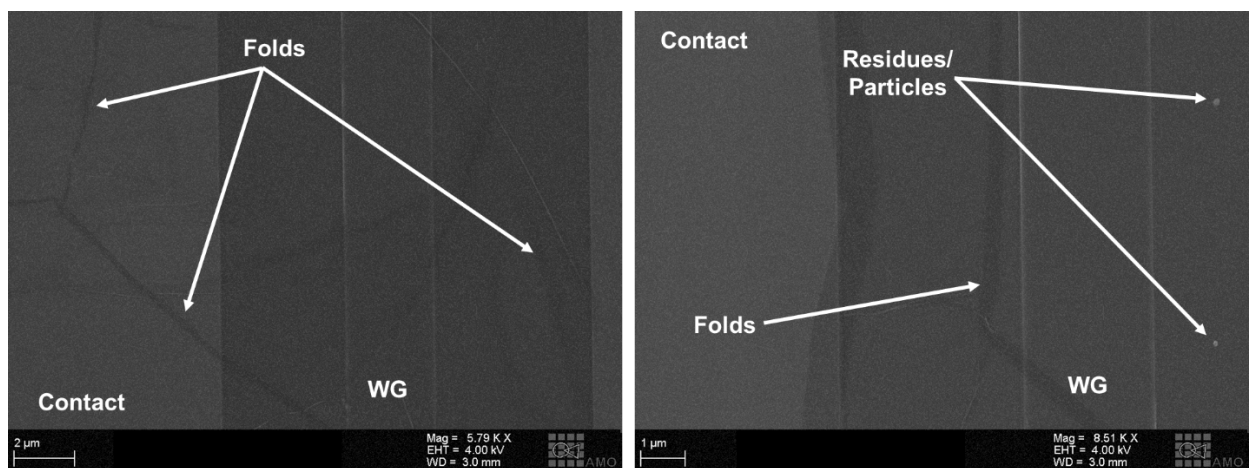

**Figure S8.** SEM images showing defects in the graphene, such as folds or residues/particles.

ACKNOWLEDGMENTS

This work has received funding from the European Union's Horizon 2020 research and innovation programme under grant agreement No 825272 (ULISSES) and No 101017186 (AEOLUS).